\documentstyle [12pt,epsf]{article}

\input{epsf}
\begin{document}
\begin{titlepage}
\begin{center}

 \vspace{-0.7in}

{\large \bf Stochastic Quantization of Topological Field Theory:\\
Generalized Langevin Equation with Memory Kernel}\\
 \vspace{.3in}{\large\em G. Menezes\,\footnotemark[1]
and N.~F.~Svaiter
\footnotemark[2]}\\
Centro Brasileiro de Pesquisas F\'{\i}sicas\,-CBPF,\\
 Rua Dr. Xavier Sigaud 150,\\
 Rio de Janeiro, RJ, 22290-180, Brazil \\

\subsection*{\\Abstract}
\end{center}
\baselineskip .18in

We use the method of stochastic quantization in a topological
field theory defined in an Euclidean space, assuming a Langevin
equation with a memory kernel. We show that our procedure for the
Abelian Chern-Simons theory converges regardless of the nature of the Chern-Simons coefficient.\\

PACS numbers: 03.70+k, 04.62.+v

\footnotetext[1]{e-mail:\,\,gsm@cbpf.br}
\footnotetext[2]{e-mail:\,\,nfuxsvai@cbpf.br}

\end{titlepage}

\newpage\baselineskip .20
in
\section{Introduction}

In the Euclidean version of field theory, we are interested in
computing the Schwinger functions of a theory. In order to obtain
these functions, Parisi and Wu introduced the stochastic
quantization \cite{parisi}. This formalism was introduced as an
alternative quantization scheme, different from the usual
canonical and the path integral field quantization, based in the
Hamiltonian and the Lagrangian respectively. The method starts
from a classical equation of motion, but not from Hamiltonian or
Lagrangian, and consequently can be used to quantize dynamical
systems without canonical formalism and therefore it is useful in
situations where the others methods lead to difficult problems.

The main idea of the stochastic quantization is that a
$d$-dimensional quantum system is equivalent to a
$(d+1)$-dimensional classical system which undergoes random
fluctuations. Some of the most important papers in the subject can
be found in Ref. \cite{ii}. A brief introduction to the stochastic
quantization can be found in the Ref. \cite{namiki1} and Ref.
\cite{sakita}. See also the Ref. \cite{damre}.

In a previous paper \cite{gsm}, we studied the stochastic
quantization of a self-interacting scalar field theory, assuming a
non-Markovian process, modifying the Langevin equation by
introducing a memory kernel \cite{fox1} \cite{kubo} \cite{z}. We
have shown that although a system with a stationary, Gaussian,
non-Markovian Langevin equation with a memory kernel and a colored
noise converges in the asymptotic limit of the Markov parameter
$\tau$ to the equilibrium, we obtain a non-regularized theory.

In this paper we would like to continue to investigate the virtues
of this non-Markovian stochastic quantization method, now employed
in the case of a topological field theory. One of the peculiar
features within this kind of theory is the appearance of a factor
of $i$ in front of the topological action in Euclidean space.
Since the topological theory does not depend on the metric of
space-time, the path integral measure weighing remains to be
$e^{iS}$ even after the Wick rotation. Another feature of a
topological action is that it is the integral of a density which
is not bounded from below in Euclidean space. So, if one attempts
to use a Markovian Langevin equation with a white noise to
quantize this theory, one will find serious problems if the factor
of $i$ is ignored. This Langevin equation will not tend to any
equilibrium in the large $\tau$ limit. So, in this sense, the use
of a Langevin equation with a complex action \cite{klauder}
becomes essential for stochastically quantizing a topological
action \cite {baulieu1} \cite{baulieu2} \cite{tzani}.

There is, in the literature, an approach to solve the above
mentioned convergence problem. Studying the purely topological
Chern-Simons theory, Ferrari $et\, al$, introduced a non-trivial
kernel in the Langevin equation \cite {ferrari}. On the other way,
Wu $et\, al$ \cite{wu} showed that the Langevin equation for a
Maxwell-Chern-Simons theory converges to the usual equilibrium
result without the need to introduce such kernel. Their method,
however, only works in the case where the Chern-Simons coefficient
is real.

We show in this paper that, if one uses a non-Markovian Langevin
equation with a colored random noise, this convergence problem may
be solved in a different way. We will apply this approach to
three-dimensional abelian Chern-Simons theory and prove that we
obtain convergence towards equilibrium even with an imaginary
Chern-Simons coefficient. To simplify the calculations we assume
the units to be such that $\hbar=c=1$.

\section{Stochastic quantization of
abelian Chern-Simons theory}\

Let us consider the following action for the three-dimensional
Maxwell-Chern-Simons theory, in Euclidean space:
\begin{equation}
S = \int d^{3}x\, \biggl(\frac{1}{4 \varepsilon^2}
A_{\mu}(x)(-\Delta \delta_{\mu \nu} +
\partial_{\mu}\partial_{\nu})A_{\nu}(x)- i \frac{\kappa}{8\pi}
\epsilon_{\mu \nu \rho}A_{\mu}(x)\partial_{\nu}A_{\rho}(x)\biggr),
\label{8}
\end{equation}
where $\Delta$ is the three-dimensional Laplace operator. At the
end of our calculations we set $\varepsilon\rightarrow\infty$ to
obtain the results for the purely topological theory, as discussed
in Ref. \cite{wu}. Notice the factor of $i$ in front of the
topological term, as mentioned before. In order to obtain the
Schwinger functions o the theory, let us use the stochastic
quantization method. Let us introduce a non-Markovian Langevin
equation given by
\begin{equation}
\frac{\partial}{\partial\tau}A_{\mu}(\tau,x)
=-\int_{0}^{\tau}ds\,M_\Lambda(\tau-s)
\frac{\delta\,S}{\delta\,A_{\mu}(x)}|_{A_{\mu}(x)=A_{\mu}(s,\,x)}+
\eta_{\mu}(\tau,x),
 \label{24}
\end{equation}
where $M_\Lambda(\tau-s)$ is a memory kernel and $\Lambda$ is a
arbitrary parameter. We will have, from Eqs.(\ref{8}) and
(\ref{24}), in momentum space:
\begin{eqnarray}
\frac{\partial}{\partial\tau}A_{\mu}(\tau,k)
&=&-\frac{k^2}{\varepsilon^2}\biggl(\delta_{\mu \nu} - \frac{k_{\mu}
k_{\nu}}{k^2}\biggr)\int_{0}^{\tau}ds\,M_\Lambda(\tau-s)
A_{\nu}(s,k)+ \nonumber\\
&& - \frac{\kappa}{4\pi}\epsilon_{\mu \nu
\rho}k_{\rho}\int_{0}^{\tau}ds\,M_\Lambda(\tau-s) A_{\nu}(s,k)+
\eta_{\mu}(\tau,k)
 \label{25}
\end{eqnarray}
where the stochastic field $\eta_{\mu}(\tau,k)$ satisfies the
modified Einstein relations:
\begin{equation}
\langle\,\eta_{\mu}(\tau,k)\,\rangle _{\eta}= 0
\end{equation}
and also
\begin{equation}
\langle\,\eta_{\mu}(\tau,k)\,\eta_{\nu}(\tau',k')\,\rangle
_{\eta}=2
\delta_{\mu\nu}M_{\Lambda}(|\tau-\tau'|)\,\delta^{d}(k+k').
\label{E}
\end{equation}
For the initial condition $A_{\mu}(\tau,k)|_{\tau=0}=0$, it is
easy to see that the solution of the Eq.(\ref{25}) is given by:
\begin{equation}
A_{\mu}(\tau,k)= \int_{0}^{\infty}d\tau'\,G_{\mu
\nu}(k;\tau-\tau')\eta_{\nu}(\tau',k), \label{26}
\end{equation}
where we introduced the retarded Green function $G_{\mu
\nu}(k,\tau)$, which satisfies:
\begin{eqnarray}
\frac{\partial}{\partial\tau}\,G_{\mu \nu}(k,\tau)
&=&-\frac{k^2}{\varepsilon^2}\biggl(\delta_{\mu \rho} -
\frac{k_{\mu}
k_{\rho}}{k^2}\biggr)\int_{0}^{\tau}ds\,M_\Lambda(\tau-s)\,
G_{\rho \nu}(k,s)+ \nonumber\\
&& - \frac{\kappa}{4\pi}\epsilon_{\mu \rho
\sigma}k_{\sigma}\int_{0}^{\tau}ds\,M_\Lambda(\tau-s)\,G_{\rho
\nu}(k,s)+ \delta_{\mu \nu}\,\delta(\tau),
 \label{27}
\end{eqnarray}
for $\tau > 0$ and $G_{\mu \nu}(k,\tau) = 0$ for $\tau < 0$.

To proceed the calculations, let us introduce the Laplace
transform of the Eq.(\ref{27}):
\begin{eqnarray}
z\,G_{\mu \nu}(k,z)
&=&-\frac{k^2}{\varepsilon^2}\biggl(\delta_{\mu \rho} -
\frac{k_{\mu} k_{\rho}}{k^2}\biggr)M_\Lambda(z)\,
G_{\rho \nu}(k,z)+ \nonumber\\
&& - \frac{\kappa}{4\pi}\epsilon_{\mu \rho \sigma}k_{\sigma}
M_\Lambda(z)\,G_{\rho \nu}(k,z)+ \delta_{\mu \nu},
 \label{28}
\end{eqnarray}
where:
\begin{equation}
M_\Lambda(z)=\int_{0}^{\infty}d\tau\,M_{\Lambda}(\tau)\,e^{-z\tau}.
\end{equation}
For the result without memory (or, formally, when
$M_{\Lambda}(\tau)\rightarrow\delta(\tau)$), we have, from
Eq.(\ref{27}):
\begin{eqnarray}
\frac{\partial}{\partial\tau}G_{\mu \nu}(k,\tau)
&=&-\frac{k^2}{\varepsilon^2}\biggl(\delta_{\mu \rho} -
\frac{k_{\mu}
k_{\rho}}{k^2}\biggr)G_{\rho \nu}(k,\tau)+ \nonumber\\
&& - \frac{\kappa}{4\pi}\epsilon_{\mu \rho \sigma}k_{\sigma}G_{\rho
\nu}(k,\tau)+ \delta_{\mu \nu}\delta(\tau),
 \label{29}
\end{eqnarray}
whose Laplace transform reads:
\begin{eqnarray}
z\,G_{\mu \nu}(k,z)
&=&-\frac{k^2}{\varepsilon^2}\biggl(\delta_{\mu
\rho} - \frac{k_{\mu} k_{\rho}}{k^2}\biggr)G_{\rho \nu}(k,z)+ \nonumber\\
&& - \frac{\kappa}{4\pi}\epsilon_{\mu \rho \sigma}k_{\sigma}G_{\rho
\nu}(k,z)+ \delta_{\mu \nu}.
 \label{30}
\end{eqnarray}
Note the similarity between Eqs.(\ref{28}) and (\ref{30}). The
solution to Eq.(\ref{29}) is given by \cite{wu}
\begin{eqnarray}
G_{\mu \nu}(k,\tau) &=& \frac{k_{\mu}
k_{\nu}}{k^2}+\Biggl(\biggl(\delta_{\mu \nu} - \frac{k_{\mu}
k_{\nu}}{k^2}\biggr)\cos\biggl(\frac{\kappa}{4\pi}k\tau\biggr) + \nonumber\\
&& - \epsilon_{\mu \nu
\sigma}\frac{k_{\sigma}}{k}\sin\biggl(\frac{\kappa}{4\pi}k\tau\biggr)\Biggr)
\exp\biggl(\frac{-k^2}{\varepsilon^2}\tau\biggr),
 \label{31}
\end{eqnarray}
whose Laplace transform is:
\begin{equation}
G_{\mu \nu}(k,z) = \frac{k_{\mu} k_{\nu}}{k^2}\frac{1}{z}+
\frac{\Biggl(\biggl(\delta_{\mu \nu} - \frac{k_{\mu}
k_{\nu}}{k^2}\biggr)z - \epsilon_{\mu \nu
\sigma}k_{\sigma}\biggl(\frac{\kappa}{4\pi}\biggr)\Biggr)}{\biggl(z+\frac{k^2}{\varepsilon^2}\biggr)^2
+\biggl(\frac{\kappa}{4\pi}\biggr)^2k^2}.
 \label{32}
\end{equation}
Comparing Eqs.(\ref{28}) and (\ref{30}), it is trivial to obtain
the analog of Eq.(\ref{32}) with memory:
\begin{equation}
G_{\mu \nu}(k,z) = \frac{k_{\mu} k_{\nu}}{k^2}\frac{1}{z}+
\frac{\Biggl(\biggl(\delta_{\mu \nu} - \frac{k_{\mu}
k_{\nu}}{k^2}\biggr)z - \epsilon_{\mu \nu
\sigma}k_{\sigma}\biggl(\frac{\kappa'}{4\pi}\biggr)\Biggr)}{\biggl(z+\frac{k^2}{\varepsilon'^2}\biggr)^2
+\biggl(\frac{\kappa'}{4\pi}\biggr)^2k^2},
 \label{33g}
\end{equation}
where:
\begin{equation}
\frac{1}{\varepsilon'^2}\equiv\frac{M_\Lambda(z)}{\varepsilon^2}
\end{equation}
and
\begin{equation}
\kappa'\equiv\kappa M_\Lambda(z).
\end{equation}
In the appendix $A$, we derive in detail the inverse Laplace
transform of Eq.(\ref{33g}). It is given by:
\begin{equation}
G_{\mu\nu}(k,\tau) = \Biggl(\frac {k_{\mu}k_{\nu}}{k^2} +
g_{\mu\nu}G_1(k,\tau) + \tilde{g}_{\mu\nu}G_2(k,\tau)
\Biggr)\theta(\tau), \label{g}
\end{equation}
where the quantities $G_i(k,\tau)$, $i=1,2$, $g_{\mu\nu}$ and
$\tilde{g}_{\mu\nu}$ are defined in the appendix A. We see that
our $G_{\mu\nu}(k,\tau)$ does not approach zero as
$\tau\rightarrow\infty$. The reason of such behavior is the
presence of the longitudinal term $\frac {k_{\mu}k_{\nu}}{k^2}$,
which is common in the stochastic quantization of all gauge
theories without gauge fixing and can be eliminated by a suitable
stochastic gauge fixing. In spite of this, the presence of this
term will not give any contribution to gauge invariant quantities.

After this discussion, we are able to present the two-point
correlation function. We have that $D_{\mu\nu}(k;\tau,\tau')$ is
given by
\begin{eqnarray}
D_{\mu\nu}(k;\tau,\tau')&\equiv&
\langle\,A_{\mu}(\tau,k)\,A_{\nu}(\tau',k')\,\rangle _{\eta}=
\nonumber\\
&&= \delta^{d}(k+k')\int_{0}^{\infty} ds\int_{0}^{\infty}
ds'\,G_{\mu\kappa}(k,\tau-s)\,G_{\lambda\nu}(k,\tau'-s')\,
\langle\,\eta_{\kappa}(s,k)\,\eta_{\lambda}(s',k')\,\rangle
_{\eta}
\nonumber\\
&& = 2 \delta^{d}(k+k')\int_{0}^{\infty} ds\int_{0}^{\infty}
ds'\,G_{\mu\lambda}(k,\tau-s)\,G_{\lambda\nu}(k,\tau'-s')\,
M_\Lambda(\mid s-s' \mid). \label{80}
\end{eqnarray}

So, inserting Eq.(\ref{g}) in the above equation and splitting the
result in five different contributions yields:
\begin{equation}
D_{\mu\nu}(k;\tau,\tau') = 2
\delta^{d}(k+k')\biggl(J_1+J_2+J_3+J_4+J_5 \biggr)
\end{equation}
where:
\begin{equation}
J_1\equiv\int_{0}^{\tau} ds\int_{0}^{\tau'} ds'\,\frac
{k_{\mu}k_{\nu}}{k^2}M_\Lambda(\mid s-s' \mid),\label{81}
\end{equation}
\begin{equation}
J_2\equiv\int_{0}^{\tau} ds\int_{0}^{\tau'}
ds'\,g_{\mu\lambda}g_{\lambda\nu}G_{1}(k;\tau-s)G_{1}(k;\tau'-s')M_\Lambda(\mid
s-s' \mid),\label{82}
\end{equation}
\begin{equation}
J_3\equiv\int_{0}^{\tau} ds\int_{0}^{\tau'}
ds'\,\tilde{g}_{\mu\lambda}\tilde{g}_{\lambda\nu}G_{2}(k;\tau-s)G_{2}(k;\tau'-s')M_\Lambda(\mid
s-s' \mid),\label{83}
\end{equation}
\begin{equation}
J_4\equiv\int_{0}^{\tau} ds\int_{0}^{\tau'}
ds'\,g_{\mu\lambda}\tilde{g}_{\lambda\nu}G_{1}(k;\tau-s)G_{2}(k;\tau'-s')M_\Lambda(\mid
s-s' \mid),\label{84}
\end{equation}
and finally
\begin{equation}
J_5\equiv\int_{0}^{\tau} ds\int_{0}^{\tau'}
ds'\,\tilde{g}_{\mu\lambda}g_{\lambda\nu}G_{2}(k;\tau-s)G_{1}(k;\tau'-s')M_\Lambda(\mid
s-s' \mid).\label{85}
\end{equation}

We can solve these equations by ordering the fictitious times $s$
and $s'$, $s > s'$ for instance, and solving the integrals in $s$
($s'$) in the interval $[0,t]$ ($[0,s]$). We obtain for $J_1$, in
the limit $\tau\rightarrow\infty$,
\begin{equation}
J_1=\frac{1}{2}\frac
{k_{\mu}k_{\nu}}{k^2}\biggl(\tau-\frac{1}{\Lambda^2}\biggr).
\label{86}
\end{equation}
The integrals $J_2$ and $J_3$ can be solved by analogy with the
scalar case \cite{gsm}. Making the following replacements:
\begin{equation}
(k^2 + m^2)_1\rightarrow
\frac{\alpha}{\Lambda^2}(1-\Lambda^4)+\Lambda^2 y_1 +
\frac{(\alpha^2-y_{1}^2)}{\Lambda^2},\label{87}
\end{equation}
\begin{equation}
(k^2 + m^2)_2\rightarrow
\frac{\alpha}{\Lambda^2}(1+\Lambda^4)-\Lambda^2 y_1 -
\frac{(\alpha^2-y_{1}^2)}{\Lambda^2},\label{87}
\end{equation}
where the subscript $1$ ($2$) stands for the $G_1$ ($G_2$) case
(see the appendix A), we will have, in the asymptotic limit
$\tau\rightarrow\infty$ that
\begin{eqnarray}
J_2 &=&\Biggl(\frac{\alpha}{\Lambda^2}(1-\Lambda^4)+\Lambda^2 y_1
+
\frac{(\alpha^2-y_{1}^2)}{\Lambda^2}\Biggr)^{-1}\Biggr[\frac{\Lambda^2}{(\sigma\gamma)^2}
\biggl(\frac{\Lambda^4}{4}+\frac{(\sigma+\gamma)^2}{4}\biggr)
\biggl(\frac{\kappa}{4\pi}\biggr)\epsilon_{\mu\nu\rho}k_{\rho}+\nonumber\\
&&+
(\sigma\gamma)^{-2}\Biggl(\biggl(\frac{\Lambda^4}{4}+\frac{(\sigma+\gamma)^2}{4}
\biggr)^2-\frac{\Lambda^4}{4}k^2\biggl(\frac{\kappa}{4\pi}\biggr)^2\Biggr)\biggl(\delta_{\mu
\nu} - \frac{k_{\mu} k_{\nu}}{k^2}\biggr)\Biggl], \label{88}
\end{eqnarray}
and
\begin{eqnarray}
J_3 &=&\Biggl(\frac{\alpha}{\Lambda^2}(1+\Lambda^4)-\Lambda^2 y_1
-
\frac{(\alpha^2-y_{1}^2)}{\Lambda^2}\Biggr)^{-1}\Biggr[-\frac{\Lambda^2}{(\sigma\gamma)^2}
\biggl(\frac{\Lambda^4}{4}+\frac{(\sigma-\gamma)^2}{4}\biggr)
\biggl(\frac{\kappa}{4\pi}\biggr)\epsilon_{\mu\nu\rho}k_{\rho}+\nonumber\\
&&+
(\sigma\gamma)^{-2}\Biggl(\biggl(\frac{\Lambda^4}{4}+\frac{(\sigma-\gamma)^2}{4}
\biggr)^2-\frac{\Lambda^4}{4}k^2\biggl(\frac{\kappa}{4\pi}\biggr)^2\Biggr)\biggl(\delta_{\mu
\nu} - \frac{k_{\mu} k_{\nu}}{k^2}\biggr)\Biggl]. \label{89}
\end{eqnarray}
The remaining integrals $J_4$ and $J_5$ can be solved without any
further complications \cite{grads}. Again, in the asymptotic limit
$\tau\rightarrow\infty$, we obtain:
\begin{eqnarray}
J_4 + J_5 &&= \nonumber\\
&&\frac{f(\Lambda,\sigma,\gamma)}{g(\Lambda,\sigma,\gamma)}\Biggl[-\frac{\Lambda^2}{(2\sigma\gamma)}
\biggl(\frac{\kappa}{4\pi}\biggr)\epsilon_{\mu\nu\rho}k_{\rho}+
\nonumber\\
&&
+(\sigma\gamma)^{-2}\Biggl(\biggl(\frac{\Lambda^4}{4}+\frac{(\sigma+\gamma)^2}{4}\biggr)
\biggl(\frac{\Lambda^4}{4}+\frac{(\sigma-\gamma)^2}{4}\biggr)+
\frac{\,\Lambda^4}{4}k^2\biggl(\frac{\kappa}{4\pi}\biggr)^2\Biggr)\biggl(\delta_{\mu
\nu} - \frac{k_{\mu} k_{\nu}}{k^2}\biggr)\Biggr], \label{90}
\end{eqnarray}
where:
\begin{eqnarray}
f(\Lambda,\sigma,\gamma)&\equiv&
153\Lambda^{14}+\Lambda^{10}\biggl[18(\sigma+\gamma)^2+17(\sigma-\gamma)^2+9\sigma\gamma\biggr]
+\Lambda^6\biggl[(\sigma+\gamma)^4+(\sigma-\gamma)^4+\nonumber\\
&&+\frac{\sigma\gamma}{2}
\biggl((\sigma+\gamma)^2+(\sigma-\gamma)^2\biggr)-\frac{9}{2}\sigma\gamma\biggl((\sigma+\gamma)^2
-\sigma\gamma(1-2\sigma\gamma)\biggr)\biggr]+\nonumber\\
&&+\Lambda^2\sigma\gamma\biggl[\frac{(\sigma+\gamma)^2}{2}-
\sigma\gamma(\sigma+\gamma)^2-\frac{1}{2}(\sigma+\gamma)^2(\sigma-\gamma)^2\biggr],
\label{91}
\end{eqnarray}
and
\begin{equation}
g(\Lambda,\sigma,\gamma)\equiv\biggl(9\Lambda^4+(\sigma-\gamma)^2\biggr)\biggl(9\Lambda^4+(\sigma+\gamma)^2\biggr)
(\Lambda^4+\sigma^2)(\Lambda^4+\gamma^2). \label{92}
\end{equation}
As mentioned before, the linearly divergent longitudinal term,
found in Eq.(\ref{86}), can be eliminated by a stochastic gauge
fixing. Now, taking the limit $\varepsilon\rightarrow\infty$, it
is easy to see that the contribution $J_2+J_3$ vanishes
identically. Then, finally, we obtain, for the purely topological
two-point correlation function:
\begin{eqnarray}
D_{\mu\nu} (k;\tau,\tau')&=& 2
\delta^{d}(k+k')\Biggl[\frac{1}{2}\frac
{k_{\mu}k_{\nu}}{k^2}\biggl(\tau-\frac{1}{\Lambda^2}\biggr)+
\frac{f'(\Lambda,\sigma,\gamma)}{g'(\Lambda,\sigma,\gamma)}
\Biggl(-\frac{\Lambda^2}{2Q(y'_1)}\biggl(\frac{\kappa}{4\pi}\biggr)
\epsilon_{\mu\nu\rho}k_{\rho}+\nonumber\\
&&+\biggl(\frac{\beta'}{Q^2(y'_1)}-1 \biggr)\biggl(\delta_{\mu
\nu} - \frac{k_{\mu} k_{\nu}}{k^2}\biggr)\Biggr)\Biggr],
\label{93}
\end{eqnarray}
where $\beta'=\beta|_{\varepsilon\rightarrow\infty}=
\frac{\Lambda^4}{4}k^2(\frac{\kappa}{4\pi})$,
$y'_1=y_1|_{\varepsilon\rightarrow\infty}$ and:
\begin{eqnarray}
f'(\Lambda,\sigma,\gamma)&=& f|_{\varepsilon\rightarrow\infty} =
120\Lambda^{14}+\frac{19}{2}\Lambda^{10}Q(y'_1)+9\Lambda^6Q^2(y'_1)+\nonumber\\
&&-\frac{\Lambda^6}{2}Q(y'_1)+\frac{9}{2}\Lambda^4Q(y'_1)+\Lambda^2Q^2(y'_1)-
18Q^2(y'_1)+\frac{9}{2}Q(y'_1), \label{94}
\end{eqnarray}
\begin{equation}
g'(\Lambda,\sigma,\gamma)=g|_{\varepsilon\rightarrow\infty} =
64\Lambda^8Q^2(y'_1)+32\Lambda^4Q^3(y'_1)+4Q^4(y'_1), \label{95}
\end{equation}
\begin{equation}
Q(y'_1)\equiv\biggl(\Lambda^4 y'_1-(y'_{1})^2\biggr)^{1/2}.
\label{96}
\end{equation}
We see that in our last expression for the propagator remained a
term proportional to the Maxwell transversal propagator. This is a
anomalous situation, since the Maxwell contribution is absent in
the usual purely topological Chern-Simons theory. The origin of
this anomalous situation is the use of a non-Markovian Langevin
equation. To circumvent this problem and recover the usual result,
we have to make the following choice:
\begin{equation}
\beta'=Q^2(y'_1),
\end{equation}
which lead us to:
\begin{equation}
y'_1=\frac{\Lambda}{2}\, \pm\, \frac{(\Lambda^8-4\beta')^{1/2}}{2}.
\label{97}
\end{equation}
So, if we choose:
\begin{equation}
y_1=\frac{\Lambda}{2}\, \pm\, \frac{(\Lambda^8-4\beta')^{1/2}}{2}+
\frac{C}{\varepsilon^n}, \label{98}
\end{equation}
where $C$ is a real constant and $n$ is an arbitrarily large
integer number. Inserting this latter equation in Eq.(\ref{r}), we
will get a cubic equation in $C$. From the usual Galois theory of
radical solutions for polynomials \cite{birkhoff} \cite{van}, we
can always choose a real root from the three possible ones. So, in
other words, we can always choose a real constant such that the
two-point correlation function converges to a ``purely
topological" term, with some minor differences from the usual one.
We notice as well that our approach still works when $\kappa$ is
purely imaginary (which is mathematically analogous to writing
$A_{\mu}=A'_{\mu}+iA'_{\mu}$, where $A'_{\mu}$ is real, and taking
the real part of the Langevin equation (\ref{25}) in coordinate
space).

\section{Conclusions}

In this paper we discussed the stochastic quantization for Maxwell
Chern-Simons theory using a non-Markovian Langevin equation and
examined the field theory that appears in the asymptotic limit of
this non-Markovian process.

This paper is the second one of a program where it is investigated
the possibility that the Parisi-Wu quantization method can be
extended assuming a Langevin equation with a memory kernel with
the modified Einstein relations. To make sure that this
modification can be used, one must first check that the system
evolves to the equilibrium in the asymptotic limit. Second we have
to show that converges to the correct equilibrium distribution. We
proved that although the system evolves to equilibrium, in the
propagator remained a term proportional to the Maxwell transversal
propagator. This is a anomalous situation, since the Maxwell
contribution is absent in the usual purely topological
Chern-Simons theory. To circumvent this problem and recover the
usual result, we have imposed a constraint in the parameters of
our theory.

\section{Acknowledgements}

 This paper was supported by the Conselho Nacional de
Desenvolvimento Cientifico e Tecnol{\'o}gico of Brazil (CNPq).

\begin{appendix}
\makeatletter \@addtoreset{equation}{section} \makeatother
\renewcommand{\theequation}{\thesection.\arabic{equation}}

\section{Appendix}
In this appendix, we derive the retarded Green function for the
diffusion problem $G_{\mu\nu}(k,\tau)$. Expanding the denominator
in Eq.(\ref{33}), given by
\begin{equation}
G_{\mu \nu}(k,z) = \frac{k_{\mu} k_{\nu}}{k^2}\frac{1}{z}+
\frac{\Biggl(\biggl(\delta_{\mu \nu} - \frac{k_{\mu}
k_{\nu}}{k^2}\biggr)z - \epsilon_{\mu \nu
\sigma}k_{\sigma}\biggl(\frac{\kappa'}{4\pi}\biggr)\Biggr)}{\biggl(z+\frac{k^2}{\varepsilon'^2}\biggr)^2
+\biggl(\frac{\kappa'}{4\pi}\biggr)^2k^2},
 \label{33}
\end{equation}
we have:
\begin{equation}
G_{\mu \nu}(k,z) =\frac{k_{\mu}
k_{\nu}}{k^2}\frac{1}{z}+\biggl(\delta_{\mu \nu} - \frac{k_{\mu}
k_{\nu}}{k^2}\biggr)I_{1}(z) - \epsilon_{\mu \nu
\sigma}k_{\sigma}\biggl(\frac{\kappa}{4\pi}\biggr)I_{2}(z),
\label{34}
\end{equation}
where:
\begin{equation}
I_{1}(z)\equiv\frac{z}{P(z)}, \label{35}
\end{equation}
\begin{equation}
I_{2}(z)\equiv\frac{M_\Lambda(z)}{P(z)},\label{36}
\end{equation}
and:
\begin{equation}
P(z)\equiv z^2 + 2\frac{k^2}{\varepsilon^2}M_\Lambda(z)z +
\frac{k^4}{\varepsilon^4}M_\Lambda^2(z) +
\biggl(\frac{\kappa}{4\pi}\biggr)^2k^2M_\Lambda^2(z). \label{37}
\end{equation}
Using the following exponential representation for the memory
kernel $M_{\Lambda}(\tau)$:
\begin{equation}
M^n_\Lambda(\tau)=\frac{1}{2n!}\Lambda^2(\Lambda^2\mid\tau\mid)^n
\exp\bigl(-\Lambda^2\mid\tau\mid\bigr), \label{38}
\end{equation}
where $\Lambda$ is a parameter, we will have, for the case $n=0$:
\begin{equation}
I_{1}(z)=\frac{z^{3}+ 2\Lambda^2 z^2 + \Lambda^4 z}{\Omega(z)},
\label{39}
\end{equation}
\begin{equation}
I_{2}(z)=\frac{\frac{\Lambda^2}{2}z +
\frac{\Lambda^4}{2}}{\Omega(z)},\label{40}
\end{equation}
and:
\begin{equation}
\Omega(z)\equiv z^4 + 2\Lambda^2 z^3 + (\Lambda^4 + \alpha)z^2 +
\alpha\Lambda^2z + \beta, \label{41}
\end{equation}
where:
\begin{equation}
\alpha\equiv\frac{k^2\Lambda^2}{\varepsilon^2}, \label{42}
\end{equation}
and
\begin{equation}
\beta\equiv\Biggl(\frac{k^4}{\varepsilon^4}+\biggl(\frac{\kappa}{4\pi}\biggr)^2k^2
\Biggr)\frac{ \Lambda^4}{4}. \label{43}
\end{equation}

In order to get the inverse Laplace transform of Eq.(\ref{34}), we
must seek for the solutions of the quartic equation $\Omega(z)=0$.
As it is well known, a general quartic equation is a fourth-order
polynomial equation of the form:
\begin{equation}
z^4 + a_3 z^3 + a_2 z^2 + a_1 z + a_0 = 0. \label{44}
\end{equation}
Using the familiar algebraic technique developed by Ferrari and
Cardano \cite{abram}, it is easy to show that the roots of
Eq.(\ref{44}) are given by:
\begin{equation}
z_1 = -\frac{1}{4} a_3 + \frac{1}{2} R + \frac{1}{2} D, \label{45}
\end{equation}
\begin{equation}
z_2 = -\frac{1}{4} a_3 + \frac{1}{2} R - \frac{1}{2} D, \label{46}
\end{equation}
\begin{equation}
z_3 = -\frac{1}{4} a_3 - \frac{1}{2} R + \frac{1}{2} E, \label{47}
\end{equation}
\begin{equation}
z_4 = -\frac{1}{4} a_3 - \frac{1}{2} R - \frac{1}{2} E, \label{48}
\end{equation}
where:
\begin{equation}
R\equiv\biggl(\frac{1}{4} a_{3}^2 - a_2 + y_1\biggr)^{1/2},
\label{49}
\end{equation}
\begin{equation}
D\equiv \left\{ \biggl(F(R)+G\biggr)^{1/2} \hfill\hbox{for
$R\neq0$} \atop \biggl(F(0)+H\biggr)^{1/2} \quad \hbox{for
$R=0,$}\right. \label{50}
\end{equation}
\begin{equation}
E\equiv \left\{ \biggl(F(R)-G\biggr)^{1/2} \hfill\hbox{for
$R\neq0$} \atop \biggl(F(0)-H\biggr)^{1/2} \quad \hbox{for
$R=0,$}\right. \label{52}
\end{equation}
\begin{equation}
F(R)\equiv\frac{3}{4} a_{3}^2 - R^2 - 2a_2, \label{54}
\end{equation}
\begin{equation}
H\equiv2\biggl(y_{1}^2-4a_0\biggr)^{1/2}, \label{55}
\end{equation}
\begin{equation}
G\equiv\frac{1}{4}(4 a_3a_2-8a_1-a_{3}^3)R^{-1}, \label{56}
\end{equation}
and $y_1$ is a real root of the following cubic equation:
\begin{equation}
y^3 - a_2 y^2 + (a_1a_3 - 4a_0)y + (4a_2a_0 - a_{1}^2 -
a_{3}^2a_0)=0. \label{r}
\end{equation}
Therefore, the inverse Laplace transform of $I_1(z)$ and $I_2(z)$
reads:
\begin{eqnarray}
I_1(\tau) &=& \frac{z_{1}^3 + 2\Lambda^2z_{1}^2+\Lambda^4z_1}{(z_1
-
z_2)(z_1 - z_3)(z_1 - z_4)} e^{z_1 \tau} + \nonumber\\
 &&+\frac{z_{2}^3 + 2\Lambda^2z_{2}^2+\Lambda^4z_2}{(z_2 - z_1)(z_2 - z_3)(z_2 - z_4)}
e^{z_2 \tau} + \nonumber\\
 &&+\frac{z_{3}^3 + 2\Lambda^2z_{3}^2+\Lambda^4z_3}{(z_3 - z_1)(z_3 - z_2)(z_3 - z_4)}
e^{z_3 \tau} + \nonumber\\
&&+\frac{z_{4}^3 + 2\Lambda^2z_{4}^2+\Lambda^4z_4}{(z_4 - z_1)(z_4
- z_2)(z_4 - z_3)} e^{z_4 \tau} , \label{57}
\end{eqnarray}
and
\begin{eqnarray}
I_2(\tau) &=& \frac{\frac{\Lambda^2}{2}z_1 +
\frac{\Lambda^4}{2}}{(z_1 -
z_2)(z_1 - z_3)(z_1 - z_4)} e^{z_1 \tau} + \nonumber\\
 &&+\frac{\frac{\Lambda^2}{2}z_2 +
\frac{\Lambda^4}{2}}{(z_2 - z_1)(z_2 - z_3)(z_2 - z_4)}
e^{z_2 \tau} + \nonumber\\
 &&+\frac{\frac{\Lambda^2}{2}z_3 +
\frac{\Lambda^4}{2}}{(z_3 - z_1)(z_3 - z_2)(z_3 - z_4)}
e^{z_3 \tau} + \nonumber\\
&&+\frac{\frac{\Lambda^2}{2}z_4 + \frac{\Lambda^4}{2}}{(z_4 -
z_1)(z_4 - z_2)(z_4 - z_3)} e^{z_4 \tau} . \label{58}
\end{eqnarray}

Now, let us study a simple convergence criterium in order that
$G_{\mu\nu}(k,\tau)\rightarrow0$ as the Markov paramter goes to
infinity, i.e., $\tau\rightarrow\infty$. In this situation, the
system converges to an equilibrium. Comparing the polynomial
$\Omega(z)$ with expression Eq.(\ref{44}), it is trivial to make
the following identifications: $a_0 = \beta$, $a_1 =
\alpha\Lambda^2$, $a_2 = \alpha + \Lambda^4$ and, finally, $a_3 =
2\Lambda^2$.

For convenience, let us assume that $R$, defined by Eq.(\ref{49}),
does not vanish. To proceed with the calculations, let us
introduce the following real quantities $\sigma$ and $\gamma$
defined respectively by
\begin{equation}
\sigma\equiv\biggl(a_2 - \frac{1}{4}a_{3}^2 - y_1\biggr)^{1/2}
=(\alpha- y_1)^{1/2} \label{63}
\end{equation}
and
\begin{equation}
\gamma\equiv\biggl(a_2 + y_1 - \frac{1}{2}a_{3}^2\biggr)^{1/2} =
(\alpha + y_1 - \Lambda^4 )^{1/2}, \label{64}
\end{equation}
where we used the identifications $a_2 = \alpha + \Lambda^4$ and
$a_3 = 2\Lambda^2$. Then, we shall have:
\begin{equation}
R = i\sigma, \label{65}
\end{equation}
and
\begin{equation}
E = i\gamma. \label{66}
\end{equation}
So, with the above identifications, it is easy to see to prove
that $G$, defined by Eq.(\ref{56}), vanishes identically.
Therefore, we will have, from Eq.(\ref{50}) and Eq.(\ref{52}),
that $D=E$. We also see that:
\begin{equation}
\sigma^2 + \gamma^2 = 2\alpha - \Lambda^4 > 0,
\end{equation}
which implies:
\begin{equation}
k^2 > \frac{\varepsilon^2\Lambda^2}{2}, \label{67}
\end{equation}
where we used Eq.(\ref{42}), which is a convergence criterium
similar to the massless scalar field case \cite{gsm}.

Thus, from Eqs.(\ref{45}) - (\ref{48}), Eq.(\ref{65}) and
Eq.(\ref{66}), we obtain the following solutions to $\Omega(z) =
0$:
\begin{equation}
z_1 = -\frac{\Lambda^2}{2} + \frac{1}{2} i\sigma + \frac{1}{2}
i\gamma, \label{68}
\end{equation}
\begin{equation}
z_2 = -\frac{\Lambda^2}{2} + \frac{1}{2} i\sigma - \frac{1}{2}
i\gamma, \label{69}
\end{equation}
\begin{equation}
z_3 = -\frac{\Lambda^2}{2} - \frac{1}{2} i\sigma + \frac{1}{2}
i\gamma, \label{70}
\end{equation}
\begin{equation}
z_4 = -\frac{\Lambda^2}{2} - \frac{1}{2} i\sigma- \frac{1}{2}
i\gamma. \label{71}
\end{equation}
So, from these last results, we will have, finally, for
$G_{\mu\nu}(k,\tau)$:
\begin{equation}
G_{\mu\nu}(k,\tau) = \Biggl(\frac {k_{\mu}k_{\nu}}{k^2} +
g_{\mu\nu}G_1(k,\tau) + \tilde{g}_{\mu\nu}G_2(k,\tau)
\Biggr)\theta(\tau), \label{72}
\end{equation}
where:
\begin{equation}
G_1(k,\tau)\equiv\Biggl(\frac{\Lambda^2}{(\sigma +
\gamma)}\sin\biggl(\frac{(\sigma + \gamma)}{2}\tau\biggr) +
\cos\biggl(\frac{(\sigma + \gamma)}{2}\tau \biggr)\Biggr)
e^{-\frac{\Lambda^2}{2}\tau}, \label{73}
\end{equation}
\begin{equation}
G_2(k,\tau)\equiv\Biggl(\frac{\Lambda^2}{(\sigma -
\gamma)}\sin\biggl(\frac{(\sigma - \gamma)}{2}\tau\biggr) +
\cos\biggl(\frac{(\sigma - \gamma)}{2}\tau \biggr)\Biggr)
e^{-\frac{\Lambda^2}{2}\tau}, \label{74}
\end{equation}
and $g_{\mu\nu}$ and $\tilde{g}_{\mu\nu}$ appearing in
Eq.(\ref{72}) are defined by:
\begin{equation}
g_{\mu\nu}\equiv\Pi_{\mu\nu} - h_{\mu\nu}, \label{75}
\end{equation}
\begin{equation}
\tilde{g}_{\mu\nu}\equiv h_{\mu\nu} -\tilde{\Pi}_{\mu\nu},
\label{76}
\end{equation}
with:
\begin{equation}
h_{\mu\nu}\equiv -
\frac{\Lambda^2}{2\gamma\sigma}\epsilon_{\mu\nu\rho}k_{\rho}\biggl(\frac{\kappa}{4\pi}\biggr),
\label{77}
\end{equation}
\begin{equation}
\Pi_{\mu\nu} \equiv
\frac{1}{\gamma\sigma}\biggl(\frac{\Lambda^4}{4}+\frac{(\sigma +
\gamma)^2}{4}\biggr)\biggl(\delta_{\mu \nu} - \frac{k_{\mu}
k_{\nu}}{k^2}\biggr), \label{78}
\end{equation}
and
\begin{equation}
\tilde{\Pi}_{\mu\nu} \equiv
-\frac{1}{\gamma\sigma}\biggl(\frac{\Lambda^4}{4}+\frac{(\sigma -
\gamma)^2}{4}\biggr)\biggl(\delta_{\mu \nu} - \frac{k_{\mu}
k_{\nu}}{k^2}\biggr). \label{79}
\end{equation}

\end{appendix}

\end{document}